\documentclass{article}
\usepackage{amssymb}

\usepackage{amsmath}
\usepackage{cmmib57}

\def\be{\begin{equation}}
\def\ee{\end{equation}}
\def\ba{\begin{eqnarray}}
\def\ea{\end{eqnarray}}

\begin{document}



\title{Schwarzschild-De Sitter black holes in $4+1$ dimensional bulk
}

\author{
\footnote{metin.arik@boun.edu.tr} Metin ARIK  and
\footnote{dciftci@boun.edu.tr} Dilek \c{C}\.IFTC\.I\\
Department of Physics, Bo\u{g}azi\c{c}i University, 34342 Bebek,
Istanbul TURKEY}
\maketitle
\begin{abstract}
We construct a static solution for $4+1$ dimensional bulk such
that the $3+1$
dimensional world has a linear warp factor and describes the Schwarzschild-dS%
$_{4}$ black hole. For $m=0$ this four dimensional universe and
Friedmann--Robertson--Walker universe are related with an explicit
coordinate transformation. We emphasize that for linear warp
factors the effect of bulk on the brane world shows up as the
dS$_{4}$ background which is favored by the big bang cosmology.

\end{abstract}

In the universe, there exists four fundamental interactions; weak,
strong, electromagnetic and gravitational. However, all the
interactions in nature are not unified. Gravity shows some
different aspects because of its long range and unquantizable
behavior and its importance in cosmology. To get a natural
explanation for the unification problem, the possibility that
gravity might not be fundamentaly four dimensional has been considered \cite%
{Randal1}-\cite{Antoniadis}. Basically these scenarios say that
our universe may be a brane embedded in some higher dimensional
space. All matter and gauge interactions live on the brane while
gravitational interactions are effective in the whole higher
dimensional space. Besides cosmology, black holes explore the
geometry of gravity and its quantum effects. Therefore gravity and
black holes can be related with the brane-world cosmology. In many
of the brane-world scenarios, the matter fields we observe are
trapped
on the brane and when they undergo gravitational collapse a black hole forms %
\cite{Chamblin}-\cite{Emparan}. This black hole becomes a higher
dimensional object. This higher dimensional object can recover the
usual physical properties of black holes and stars.

Most of the brane world scenarios originated from Randall--Sundrum
(R-S) models in which the four dimensional Minkowski universe is
embedded in five dimensional AdS bulk. This bulk is compatible
with a Friedmann--Robertson--Walker (FRW) brane. The most general
vacuum bulk with a
FRW brane is Schwarzschild-anti-de Sitter space-time \cite{149 Ida},\cite%
{Bewcock} where a domain wall moving in the 5 dimensional
Schwarzschild-AdS
space -time with the metric%
\begin{eqnarray}
ds^{2} &=&g_{\mu \nu }dx^{\mu }dx^{\nu } \\
&=&-f\left( r\right) dt^{2}+\frac{dr^{2}}{f\left( r\right)
}+r^{2}\left[ d\chi ^{2}+f_{k}\left( \chi \right) d\Omega
_{2}^{2}\right]  \notag
\end{eqnarray}%
is considered. Here the part in the square parenthesis is the
metric of a
unit three-dimensional sphere, plane or hyperboloid for $k=1,0,-1$ and $%
f_{1}\left( \chi \right) =\sin \chi $, $f_{0}\left( \chi \right)
=\chi ,$ and $f_{-1}\left( \chi \right) =\sinh \chi $
respectively. The solution of field equations $R_{\mu \nu
}^{5}=\Lambda ^{5}g_{\mu \nu }^{5}$ with negative cosmological
constant $\Lambda ^{5}=-\frac{4}{l^{2}}$ is given by
\begin{equation*}
f\left( r\right) =k+\frac{r^{2}}{l^{2}}-\frac{\mu }{r^{2}},
\end{equation*}%
where $l$ and $\mu $ are constants. For $\mu =0$, this reduces to
AdS$_{5}$ and $\mu \neq 0$ generates the electric part of the Weyl
Tensor. By this consideration the location of three branes in the
five dimensional bulk can be investigated.

Instead of these theories we have considered a four dimensional
curved space-time with positive cosmological constant embedded in
five dimensional flat and empty Minkowski Universe as shown in
\cite{biz}. Starting from this point of view, in this paper, we
consider a usual four dimensional Schwarzschild metric embedded in
five dimensional Ricci flat universe. Our metric describes a black
hole on a domain wall at fixed $w.$ As such a space we take the
five dimensional analogue of Schwarzschild-de Sitter space-time in
the following form.
\begin{equation}
ds^{2}=dw^{2}+\left( 1-\frac{\left| w\right| }{w_{0}}\right)
^{2}\left[ -U\left( r\right) dt^{2}+\frac{1}{U\left( r\right)
}dr^{2}+r^{2}d\Omega _{2}^{2}\right] ,  \label{metric}
\end{equation}%
where we choose
\begin{equation}
U\left( r\right) =1-\frac{2M}{r}-\frac{r^{2}}{w_{0}^{2}},
\label{u(r)}
\end{equation}%
such that for $M=0$ the bulk is flat and for $M\neq 0$ the bulk is
Ricci flat. This metric tells us that for constant $w$ the
cosmological constant is $\Lambda _{observed}=3/w_{0}^{2}.$ It is
precisely the condition that the parameter $w_{0}$ in $U\left(
r\right)$ Eq. (\ref{u(r)}) and in the warp factor in Eq.
(\ref{metric}) are the same which causes the $4+1$ dimensional
bulk to be Ricci flat. We set up the $3+1$ branes at $w=0$ and
$w=w_{1}\leq w_{0}$. Then, on the branes this metric will be in
the standard form of the Schwarzschild metric. Here note that in
small scales the cosmological curvature is much smaller than the
Schwarzschild curvature. If we denote the perturbed horizon as
$r=2M+\varepsilon $ for small $\varepsilon $, the black hole
horizon is determined to be at $r=2M+8M^{3}/w_{0}^{2}$ where
$w_{0}\gg M $. On the other hand from the square of the Riemann
tensor one finds curvature singularities at $r=0$ and $w=\pm
w_{0}.$ However one expects that masses confined to the brane will
generate the localized gravitational fields, so that these
singularities are not physically important. Strictly
speaking the warp factor is given by $1-\left| w\right| /w_{0},$ for $%
-w_{1}\leq {w}\leq {w_{1},}$ and imposing periodicity of period
$2w_{1}$ together with $Z_{2}$ symmetry $w\rightarrow -w$.

For $M=0$ the four dimensional part in square parenthesis in Eq.
(\ref{metric})
transforms into%
\begin{equation}
ds_{4}^{2}=\left[ -d\tau ^{2}+g\left( \tau \right) ^{2}\left(
d\chi ^{2}+f_{k}\left( \chi \right) ^{2}d\Omega _{2}^{2}\right)
\right] , \label{ds4}
\end{equation}%
by the following transformations. Here $\tau $ and $g\left( \tau
\right) $ correspond to cosmic time and scale factors of FRW
universe respectively.

\begin{itemize}
\item $k=-1$%
\begin{eqnarray*}
&&r=w_{0}\sinh \left( \tau \right) \sinh \left( \chi \right) \\
&&t=\frac{w_{0}}{2}\ln \left( \frac{\cosh \left( \chi \right)
\sinh \left( \tau \right) +\cosh \left( \tau \right) }{\cosh
\left( \chi \right) \sinh
\left( \tau \right) -\cosh \left( \tau \right) }\right) \\
&&g\left( \tau \right) =w_{0}\sinh \left( \tau \right) \\
&&f_{-1}\left( \chi \right) =\sinh \left( \chi \right) .
\end{eqnarray*}

\item $k=0$%
\begin{eqnarray*}
&&r=\chi \exp \left( \tau /w_{0}\right) \\
&&t=\tau -\frac{w_{0}}{2}\left[ \chi ^{2}\exp \left( 2\tau /w_{0}\right) -1%
\right] \\
&&g\left( \tau \right) =\exp \left( \tau /w_{0}\right) \\
&&f_{0}\left( \chi \right) =\chi.
\end{eqnarray*}

\item $k=1$%
\begin{eqnarray*}
&&r=w_{0}\cosh \left( \tau \right) \sin \left( \chi \right) \\
&&t=\frac{w_{0}}{2}\ln \left( \frac{\cosh \left( \tau \right) \cos
\left( \chi \right) +\sinh \left( \tau \right) }{\cosh \left( \tau
\right) \cos
\left( \chi \right) -\sinh \left( \tau \right) }\right) \\
&&g\left( \tau \right) =w_{0}\cosh \left( \tau \right) \\
&&f_{1}\left( \chi \right) =\sin \left( \chi \right) .
\end{eqnarray*}
\end{itemize}

For the case of $M\neq 0,$ we can make the same transformation,
however it will not make our work easier. If we put $M=0$ in this
transformation we obtain the same metric as in Eq. (\ref{ds4}).
This means that, our space-time becomes FRW universe far away from
the black hole. Hence we can use the context of FRW cosmological
model. In this part we suppose that the visible
world is located at $w=w_{1}$. Therefore the visible metric tensor will be $%
g_{\mu \nu }^{\left( vis\right) }=g_{\mu \nu }\left( x^{\mu
},w=w_{1}\right) $ and for the hidden brane $g_{\mu \nu }^{\left(
hid\right) }=g_{\mu \nu
}\left( x^{\mu },w=0\right)$. The setup is similar to the scenario of \cite%
{Randal2}. At first, in order to get a realistic theory we need to
look for the localization of gravity on the brane. To do this,
taking $\left( 1-\left| w\right| /w_{0}\right) g_{\mu \nu }+h_{\mu
\nu }\left( x,w\right) ,$ we study the fluctuations of the metric.
Here $h_{\mu \nu }\left( x,w\right) $ is the perturbed metric and
its transverse traceless components represents the graviton on the
brane i.e. the conditions $\partial _{\mu }h^{\mu \nu
}=0 $ and $h_{\mu }^{\mu }=0$ must be satisfied \cite{Randal1},\cite{Randal2}%
,\cite{Brevik}-\cite{Padmanabhan}. By separation of variables we write $%
h_{\mu \nu }$ in terms of the four-dimensional mass eigenstates
$h_{\mu \nu }\left( x,w\right) =\phi _{\mu \nu }\left(
t,x^{i}\right) \psi \left( m,w\right) $, where $m$ corresponds to
four-dimensional mass. Performing a coordinate transformation from
$w$ to the variable $z$ defined by $\partial
z/\partial w=\pm \left( 1-\left| w\right| /w_{0}\right) ^{-1}$ and defining $%
u\left( z\right) $ by $\psi =\left( 1-\left| w\right|
/w_{0}\right) ^{-3/2}u\left( z\right) $ we get the wave function
as the solution of
\begin{equation}
\left( -\partial _{z}^{2}+V\left( z\right) \right) u\left(
z\right) =m^{2}u\left( z\right) ,  \label{schro}
\end{equation}%
where the potential $V\left( z\right) $ is%
\begin{equation}
V\left( z\right) =\frac{3}{4}\frac{f^{\prime 2}}{f^{2}}+\frac{3}{2}\frac{%
f^{\prime \prime }}{f}.
\end{equation}%
Here we denote the derivative with respect to $z$ by prime and $f$
is the $\
z$ dependent warp factor as addressed below. Since our brane is located at $%
w=w_{1},$ we can write the warp factor as $f\left( w\right) =1-\frac{w_{1}}{%
w_{0}}+\frac{\left| w_{1}-w\right| }{w_{0}},$ then $z$ dependence
of the warp factor will be
\begin{equation}
f\left( z\right) =\left( 1-\frac{w_{1}}{w_{0}}\right) ^{-1}\exp
\left( -\left| z\right| /w_{0}\right) ,
\end{equation}%
the transformed coordinate $z$ is given by
\begin{equation}
z=sign\left( w_{1}-w\right) \ln \left( \frac{1-\frac{w_{1}}{w_{0}}}{1-\frac{%
w_{1}}{w_{0}}+\frac{\left| w_{1}-w\right| }{w_{0}}}\right) ,
\end{equation}%
where our brane is placed at $z=0$. Then we find the potential as
\begin{equation}
V\left( z\right) =\frac{9}{4w_{0}^{2}}-\frac{3}{w_{0}}\delta
\left( z\right) .
\end{equation}%
Here the $\delta $ function in the potential allows a normalizable
bound state mode and this shows that the 5D graviton is localized
on the brane. In
general for a potential $V(z)$ which allows bound states, Eq. (\ref{schro}) has $%
m=0$ normalizable solutions iff $V\left( z\right) >0,$ as
$z\rightarrow
\infty $. In this case localization of gravity on the brane occurs. Here $%
m=0 $ mode corresponds to the gravitons on the brane.

The next point is to solve the Schrodinger Equation (\ref{schro}).
First of all, for the zero mode wave function, we find that
$u_{0}\left( z\right)
=\left( 3/2w_{0}\right) ^{1/2}\exp \left( -3w_{0}\left| z\right| /2\right) $%
. On the other hand, the massive modes are $u_{m}\left( z\right)
=A\exp \left( -\sqrt{\frac{9}{4w_{0}^{2}}-m^{2}}\left| z\right|
\right) ,$ where $A$ is the normalization factor. As explained in
\cite{Ito}, the massive modes for $0<m^{2}<9/4w_{0}^{2}$ do not
exist there and for $m^{2}\geq 9/4w_{0}^{2} $ this wave function
becomes plane wave. Then there is a mass gap $9/4w_{0}^{2}$
between the zero mode and the continuous modes.

At this stage, we look for the five dimensional Einstein tensor,
\begin{equation}
G_{MN}^{5}=-3\delta _{M}^{\mu }\delta _{N}^{\nu }\eta _{\mu \nu }\frac{1}{%
(w_{0}-\left| w\right| )}\left[ \delta \left( w\right) -\delta
\left( w-w_{1}\right) \right] ,  \label{gmn}
\end{equation}%
where $\eta _{\mu \nu }=diag\left( -1,1,1,1\right) $, Capital Latin letters $%
M,N=0,1,2,3,5$ denote full space-time and lower Greek $\mu ,\nu
=0,1,2,3$ run over the brane world. From Eq. (\ref{gmn}), it is
easily seen that there are no nongravitational fields ($T_{MN}=0$)
and no cosmological constant in the
bulk. For the solution on the brane, we introduce the boundary terms as%
\begin{equation}
V_{vis}=\frac{-24M_{5}^{3}}{w_{0}\left( 1-\frac{\left| w_{1}\right| }{w_{0}}%
\right) }\mbox{ \ \ \ \ \ and \ \ \ \
}V_{hid}=\frac{24M_{5}^{3}}{w_{0}}.\mbox{\ \ \ \ }  \label{V}
\end{equation}%
These results are similar but not the same as RS model in which
$V_{vis}$ and $V_{hid}$ are same with opposite sign. We can derive
the induced brane gravity \cite{Kofinas}, which consists of four
dimensional Einstein gravity on the brane. From the dimensionful
constants $\kappa _{5,}$ and $\kappa
_{4} $ the Planck masses $M_{5},$ $M_{4}$ are defined as%
\begin{equation}
\kappa _{5}^{2}=8\pi G_{(5)}=M_{5}^{-3},\mbox{ \ \ \ \ \ \ \ \ \ \
\ }\kappa _{4}^{2}=8\pi G_{\left( 4\right) }=M_{4}^{-2}.
\end{equation}%
The induced Einstein tensor on the brane is
\begin{equation}
\frac{M_{4}^{2}}{2}\int d^{4}x\sqrt{g^{4}}G_{\mu \nu }^{4}=-\frac{6M_{5}^{3}%
}{2}\int
d^{4}x\int_{-w_{1}}^{w_{1}}\frac{dw}{w_{0}}\sqrt{g^{5}}g_{\mu \nu
}\left( 1-\frac{\left| w\right| }{w_{0}}\right) ^{-1}\left[ \delta
\left( w\right) -\delta \left( w-w_{1}\right) \right] ,
\end{equation}%
then for the hidden brane world%
\begin{equation}
G_{\mu \nu \left( hid\right) }^{4}=-\frac{3}{w_{0}^{2}}\eta _{\mu \nu }%
\mbox{ \ \ \ and \ \ \ }M_{4}^{2}=\frac{V_{hid}w_{0}^{2}}{12},
\label{hidd}
\end{equation}%
and for the visible world%
\begin{equation}
G_{\mu \nu \left( vis\right) }^{4}=-\frac{3}{w_{0}^{2}\left( 1-\frac{w_{1}}{%
w_{0}}\right) ^{2}}\eta _{\mu \nu }\mbox{ \ \ \ and \ \ \ \ \ }M_{4}^{2}=%
\frac{V_{vis}w_{0}^{2}}{12}\left( 1-\frac{w_{1}}{w_{0}}\right)
^{2}, \label{vis}
\end{equation}%
where $M_{4}$ and $M_{5}$ have dimensions (length)$^{-1}$
\cite{Kofinas} and $w_{0}$ corresponds to the $1/k$ term in
\cite{Randal2}. The Einstein tensor on the brane is found as
exactly what we want to get. From the four dimensional part in the
square parenthesis in Eq. (\ref{metric}) we determine the same
result for the Einstein tensor or matter content. This is the
usual well known Schwarzschild space-time on the visible and
hidden world.
Therefore we get that the cosmological constant of the hidden world is $%
\Lambda _{4\left( hid\right) }=3/w_{0}^{2}$. On the other hand for
the visible brane we get the positive cosmological constant
$\Lambda _{4\left( vis\right) }=3\left( w_{0}-w_{1}\right) ^{-2}$
and the five dimensional cosmological constant $\Lambda _{5}$ is
zero everywhere. This may be interpreted as the four dimensional
cosmological constant originating from the localized energy
momentum tensor on the hidden brane affecting the visible brane
through the bulk. In Eq. (\ref{V}) the minus sign on the visible
brane tension comes from the attractive gravitational force. This
minus sign causes us to measure positive cosmological constant on
the visible world.

To find the mass parameter on the 3-brane we find the effective action as in %
\cite{Randal2}. Then we get
\begin{equation}
m=\left( 1-\frac{\left| w_{1}\right| }{w_{0}}\right) m_{0}.
\end{equation}%
Here if the part in parenthesis is of order $10^{-15}$, near the
curvature singularity $w_{1}\simeq w_{0}$ where the visible world
is located, we get the $TeV$ mass scale from the Planck mass
scale, $10^{16}TeV.$

These results have some importance in some sense:

\begin{itemize}
\item $V_{vis}=-10^{15}V_{hid}.$

\item The observed cosmological constant ($10^{-52}m^{-2}$) comes
from the effect of the hidden brane on our world. Really what we
observe as a cosmological constant $10^{-52}m^{-2}$ is the
$\Lambda _{4(hid)}.$ From this consideration we find $w_{o}$ from
the equation
\begin{equation}
10^{-52}m^{-2}=\frac{3}{w_{0}^{2}},
\end{equation}%
then $w_{1}\simeq w_{o}=10^{26}m$ is approximately just the Hubble
length. Therefore we are located at the distance of nearly Hubble
length from the hidden brane. On the other hand, the cosmological
constant of the visible world which we cannot observe is
$10^{-22}m^{-2}.$ The factor $10^{30}$ in the two cosmological
constants comes from the location of our world. We are much closer
to singularity $w_{0}$ than the hidden brane. Really there is no
cosmological constant more than one. To prevent this complication,
we do not use the name $\Lambda _{(vis)}$ anymore$,$ we call it
$G_{00}(w)$. We can
summarize these results, by the relation%
\begin{equation}
\Lambda _{(hid)}=\Lambda =G_{00}(w)\left( 1-\frac{\left| w\right| }{w_{0}}%
\right) ^{2}=10^{-52}m^{-2},
\end{equation}%
which is satisfied for all $w.$ Placing our brane at $w=w_{1}$ causes $%
G_{00}(w_{1})$ to be $10^{-22}m^{-2}.$ If we had placed our brane
at any other $w,$ $G_{00}$ on the brane would change as indicated
in this equation but $\Lambda $ would not change. $G_{00}(w_{0})$
becomes infinite on the
singularity at $w=w_{0}$. However $0\leq w\leq w_{1}<w_{0}$ and the point $%
w=w_{0}$ is outside the branes and the bulk.
\end{itemize}

One of the main motivations for this work has been the argument
that, since there is nothing except gravity in the fifth
dimension, there should also be no cosmological constant in the
fifth dimension. Gravity originates from the hidden brane and then
it is localized and interacts with the matter on the brane we
live. We cannot see the hidden brane. However we can measure its
effect as a cosmological constant $\Lambda $. Additionally the
distance from the hidden brane determines the localized mass
density of the universe.
Namely $G_{00}(w_{1})$, the mass density in the visible brane depends on $%
\Lambda $\ and varies with the location of the universe in the
bulk. This causes the energy density to be obtained as $\rho
\simeq 10^{3}kg/m^{3}$. This value is too large to be interpreted
as an average energy density for our universe. However,
interestingly enough this value is comparable to stellar energy
densities. This indicates that a model where the cosmological
constant is localized and creates localized masses may be viable.

At the end, the correction to Newton's Law is obtained as
\begin{eqnarray}
U\left( r\right) &=&-G_{N}\frac{m_{1}m_{2}}{r}-M^{-3}\int_{m_{0}}^{\infty }dm%
\frac{m_{1}m_{2}e^{-mr}}{r}\left( u_{m}\left( 0\right) \right) ^{2} \\
&=&-G_{N}\frac{m_{1}m_{2}}{r}(1+C\frac{{e^{-m_{0}r}}}{r}),  \notag
\end{eqnarray}%
where as we have explained above $m_{0}=3/2w_{0}$ and
$A^{2}=\left[ \ln \left( 1-\frac{w_{1}}{w_{0}}\right) ^{-1}\right]
^{-1},$ and then the constant $C=16\pi w_{0}\left(
1-\frac{w_{1}}{w_{0}}\right) \left( -\ln \left(
1-\frac{w_{1}}{w_{0}}\right) \right) ^{-1}\simeq $ $10^{11}m$.
This value of $C$ is severely in violation of experimental
measurements.

In conclusion, in this paper, we have presented a new brane world
black hole
solution such that the Schwarzschild-dS$_{4}$ space-time is embedded into $%
4+1$ dimensional bulk with the linear warp factor $(1-|w|/w_{0})$.
Here the visible brane is located at $w=w_{1}<w_{0}$. We have
presented the coordinate transformation between the
Schwarzschild-dS$_{4}$ and FRW
space-time for $M=0$. The same transformation can also be used for $M\neq 0$%
. Although such a transformation will result in a complicated metric, for $%
M=0$ this metric will reduce to FRW. For $M\neq 0,$ we found that
in the bulk not only the energy momentum tensor but also the
cosmological constant is zero. However the effect of the hidden
brane to our world is just the cosmological constant which we now
measure. The localized energy momentum tensor in the bulk also
serves to cause a four dimensional cosmological constant on the
brane. It was found that this cosmological constant depends on the
distance scale $w_{0}$. Finally we have derived the range of fifth
dimension from the hierarchy between the weak and Planck scales.
This model is in violation of observed values concerning
corrections to the Newtonian corrections and the matter-average
energy density of the universe.


\begin{thebibliography}{0}

\bibitem{Randal1} L. Randall, R. Sundrum, {\it Phys. Rev. Lett.} {\bf 83},
4690-4693 (1999).

\bibitem{Randal2} L. Randall, R. Sundrum, {\it Phys. Rev. Lett.} {\bf 83},
3370-3373 (1999).

\bibitem{Dvali} I. Antoniadis, N. Arkani-Hamed, S. Dimopoulos, and G. Dvali,
{\it Phys. Lett.} {\bf B436}, 257 (1998).

\bibitem{Antoniadis} I. Antoniadis, N. Arkani-Hamed, S. Dimopoulos, and G.
Dvali, {\it Phys. Rev.} {\bf D59}, 086004 (1999).

\bibitem{Chamblin} A. Chamblin, H.S. Reall, H.A. Shinkai, and T. Shiromizu,
Phys. Rev. {\bf D63}, 064015 (2001).

\bibitem{Hawking} A. Chamblin, S.W. Hawking, and H.S. Reall,
Phys. Rev. {\bf D61}, 065007 (2000).

\bibitem{Ruth} R.Gregory, R. Whisker, K. Beckwith, and C. Done, {\it
JCAP} {\bf 10}, 013 (2004).

\bibitem{Kofinas} G. Kofinas, {\it Phys. Rev.} {\bf D66}, 104028 (2002).

\bibitem{Myung} Y. S. Myung, {\it Class. Quantum Grav.}  {\bf 21}, 1279 (2004).

\bibitem{Nojiri} S. Nojiri, S. D. Odinstov, S. Ogushi, {\it Phys. Rev.} {\bf %
D65, }023521 (2002).

\bibitem{Marteens} N.K. Dadhich, R. Maartens, P. Papadopoulos, and V.
Rezania, {\it Phys. Lett.} {\bf B487}, 1 (2000).


\bibitem{Marteens2} R. Maartens, Living Rev. Relativity {\bf 7}, (2004).

\bibitem{Takayuki} H. Takayuki, and K. Gungwon, {\it Phys. Rev.} {\bf D64},
064010 (2001).

\bibitem{Casadio} R.Casadio, A. Fabbri, and L. Mazzacurati, {\it Phys. Rev.}
{\bf D65}, 084040 (2002).

\bibitem{Mazzacurati} R. Casadio, L. Mazzacurati, {\it Mod. Phys. Lett.} {\bf A18}, 651 (2003).

\bibitem{Emparan} R. Emparan, G.T. Horowitz, and R.C. Myers, {\it JHEP} {\bf 01}, 021 (2000.)

\bibitem{149 Ida} D. Ida, {\it JHEP} {\bf 09}, 014 (2000).

\bibitem{249} S. Mukohyama, T. Shiromizu, and K. Maeda, {\it Phys. Rev.} {\bf %
D62}, 024028 (2000).

\bibitem{Bewcock} P. Bewcock, C. Charmousis, and R. Gregory, Class. Quantum
Grav. {\bf 17}, 4745 (2000).

\bibitem{biz} M. Ar\i k, et al., {\it Phys. Rev.} {\bf D68}, 123503 (2003).

\bibitem{Brevik} I. Brevik, K. Ghoroku, S. D. Odintsov, M. Yahiro, {\it Phys.
Rev.} {\bf D66}, 064016 (2002).

\bibitem{Csaki} C. Csaki, J. Erlich, T. J. Hollowood, Y. Shirman, {\it Nucl.
Phys.} {\bf B581}, 309 (2000).

\bibitem{Poritz} J. F.Vazquez-Poritz, {\it JHEP} {\bf 0209}, 001 (2002.)

\bibitem{Ito} M. Ito, {\it Europhys. Lett.} {\bf 64}, 295 (2003).

\bibitem{Padmanabhan} T. Padmanabhan, S. Shankaranarayanan, {\it Phys. Rev.}
{\bf D63}, 105021 (2001).

\bibitem{Garriga} J. Garriga, T. Tanaka, {\it Phys. Rev. Lett.} {\bf 84, }%
2778 (2000).

\end{thebibliography}
\end{document}